  \documentclass[preprint]{aastex}
\usepackage{epsf}
\usepackage{epsfig}
\usepackage{amsmath}
\newcommand{\simgt}{\lower.5ex\hbox{$\; \buildrel > \over \sim \;$}}
\newcommand{\simlt}{\lower.5ex\hbox{$\; \buildrel < \over \sim \;$}}

\begin{document}
\title{Multi-wavelength emission region of $\gamma$-ray
emitting pulsars }

\author{S. Kisaka$^*$ and Y. Kojima$^{**}$}
\affil{
Department of Physics, Hiroshima University, Higashi-Hiroshima
739-8526, Japan }
\email{$^*$ kisaka@theo.phys.sci.hiroshima-u.ac.jp}
\email{$^{**}$ kojima@theo.phys.sci.hiroshima-u.ac.jp}
%

\begin{abstract}
Using the outer gap model, we investigate the emission region for 
the multi-wavelength light curve from energetic pulsars. 
We assume that $\gamma$-ray 
and non-thermal X-ray photons are emitted
from a particle acceleration region in the outer magnetosphere,
and UV/optical photons originate above that region.
We assume that $\gamma$-rays are radiated only 
by outwardly moving particles, whereas the other photons are 
produced by particles moving inward and outward.
We parameterize the altitude of the emission region as the 
deviation from the rotating dipole in vacuum 
and determine it from the observed
multi-wavelength pulse profile 
using the observationally constrained magnetic dipole inclination 
angle and viewing angle of the pulsars. 
We find that the outer gap model can 
explain the multi-wavelength pulse behavior
by a simple distribution of emissivity, and 
discuss the possibility of further improvement. 
From observational fitting, we also find a general tendency for 
the altitude of the $\gamma$-ray emission region 
to depend on the inclination angle. 
In particular, the emission region for low inclination angle is 
required to be located in very low altitude, which corresponds to 
the inner region within the last-open field line of rotating dipole in vacuum. 
This model suggests a modification of statistics about observed 
$\gamma$-ray pulsars.
Number of the sources with low inclination and viewing angles 
increases compared with previous estimate.
%
\end{abstract}
%
\section{INTRODUCTION}

  Pulsars emit over a wide range of energies from radio to 
$\gamma$-ray.
Recent observations by the {\it Fermi} Gamma-Ray Space 
Telescope of more than sixty pulsars \citep{Abcat, Pa10} have revealed 
further details of the structure of the emission region. 
The detection of the emissions in the GeV
energy range from a pulsar magnetosphere means that electrons and
positrons are accelerated to more than
$\sim 10^{12}$eV by the electric field parallel to the magnetic field,  
which arises in a depleted region of the 
Goldreich-Julian charge density \citep{GJ69}.
The light curve in the $\gamma$-ray band is an important tool 
for probing the particle acceleration and dissipation processes 
in the pulsar magnetosphere, since the maximum energy is determined
by the acceleration-radiation-reaction-limit for typical energetic pulsars.
The $\gamma$-ray emission region has therefore been 
explored by comparing theoretical models with the observed light 
curve(e.g., \citet{Wa09, VHG09, RW10}).
The pulsed emission is also detected in other energy bands 
(X-ray, ultraviolet, optical and radio) for 
some sources (e.g., \citet{T04}). 
The spectral features are non-thermal except for 
the soft X-ray range, and the light curves from a single object are, 
in general, different from one energy band to another. 
For example, profiles of the light curve in one spin period
are different in the $\gamma$- and X-ray ranges in the Vela pulsar
\citep{Abvela2}. 
The peak phase of different energy range is expected to coincide, since
  the emitting particles are related to a pair cascade process. However, the
  observation shows that the phase depends on the energy bands. This means
  that their emission regions are not the same region.
%
A complete understanding of 
light curve behavior 
in multi-wavelength bands can 
provide valuable information about the particle acceleration region.
Note that we do not discuss soft X-rays, 
which are believed to be thermal radiation from the neutron star 
surface (e.g., \citet{Ja02}).

  Possible origins of non-thermal pulsed emissions have been considered
in the polar cap \citep{DH96}, slot gap \citep{MH04}, and
outer gap \citep{CHR86} models. 
Recent {\it Fermi} observations with high $\gamma$-ray photon 
number statistics have showed that 
the phase-averaged spectrum above 200~MeV is well fitted 
by a power law plus exponential cutoff, and that 
a cutoff shape sharper than a simple exponential is 
rejected with high significance (e.g., \citet{Abvela2}).
This rules out the near-surface emission proposed in polar cap
cascade models \citep{DH96}, which would exhibit a much sharp spectral
cutoff due to magnetic pair-production attenuation.
Thus, pulsed $\gamma$-ray emission originates in the outer magnetosphere,
as considered in the outer gap model.

\citet{TCS08} 
(hereafter TCS08) considered a three-dimensional geometrical emission model
to fit the observed light curves at different energy bands.
%
%
The model is extended with some model parameters  
from the calculated gap structure in a two-dimensional meridian plane. 
By comparing the light curves of the Vela pulsar, 
they found that the X-ray emission is produced by both 
inward and outward emission from the gap region, 
and that UV/optical emission originates from secondary pairs at a higher 
altitude.
The number of light curves of pulsars observed at $\gamma$-ray 
and other energy bands is increasing thanks to {\it Fermi}, so that 
it is worthwhile to investigate whether outer gap model is applicable to 
other sources.

In this paper, we investigate the emission regions of several pulsars 
by fitting the simplified model of TCS08 to the observed 
multi-wavelength light curves. In this model, we have to specify 
the locations of the upper and lower boundaries of the gap region where the
non-corotation potential is zero. Therefore, we explicitly introduce 
the altitude of the gap region as a parameter, 
in order to fit the observational data easily.
The light curves also depend on the
dipole inclination and viewing angles.
In our method, such parameters are eliminated by other observational
data, and only the altitude is changed for the fitting. 
In the most studies, the lower boundary of the emission region is 
chosen as the surface of the last-open field lines of the 
rotating dipole(e.g. \citet{TCS08, RW10})
In this paper, however, the altitude is allowed to be in a wide range 
in order to explore the possible deviation of magnetic field-line 
structure from that of a rotating dipole in vacuum.
In Section 2, we describe the model assumptions and parameters.
In Section 3, we compare the peaks of light curves 
with those observed at multiple wavelengths,
and determine the altitude parameter. 
Our discussion is presented in Section 4, and
lastly, a summary is given in Section 5.

\section{MODEL DESCRIPTION}

The numerical method for fitting the light curve is well described 
by \citet{RW10} and \citet{BS10}; but we briefly summarize it in this section 
to explain one modification. Our model is almost the same as that used
by TCS08. However, we explicitly introduce the altitude of the 
emission region as an additional parameter.

We assume that magnetic field structure is approximately described 
by a rotating dipole with magnetic moment $\mu$.
The angular velocity is $\Omega$, and the magnetic axis is declined 
by an angle $\alpha$ from the axis of rotation (the z-axis). 
The magnetic moment changes with time $t$ as
\begin{equation}
\boldsymbol{\mu}(t) = \mu (\sin \alpha \cos \Omega t \hat{\bf x} 
+ \sin \alpha \sin 
 \Omega t\hat{\bf y} + \cos \alpha \hat{\bf z}).
\label{2.1}
\end{equation}
The magnetic field produced by the rotating dipole (e.g. \citet{J75})
can be expressed using the retarded time $t_r = t-r/c$ as
\begin{equation}
{\bf B}=-\biggl[ \frac{\boldsymbol{\mu}(t_r)}{r^3} + \frac{
 \dot{\boldsymbol{\mu}}(t_r)}{cr^2} + \frac{
 \ddot{\boldsymbol{\mu}}(t_r)}{c^2r} 
 \biggr] +
 \hat{\bf r}\biggl[ \hat{\bf r} \cdot \biggl(3\frac{\boldsymbol
 {\mu}(t_r)}{r^3} + 3\frac{ \dot{\boldsymbol{\mu}}(t_r)}{cr^2} + \frac{
 \ddot{\boldsymbol{\mu}}(t_r)}{c^2r} \biggr) \biggr],
\label{2.2}
\end{equation}
where $r$ is radial distance from the center of the star, and 
a dot denotes a derivative with respect to $t$.

We assume that radiation direction aligns with magnetic field
in a frame rotating with angular 
velocity $\Omega$ in which
the electric field vanishes.
Physically, this means that the magnetosphere is filled with a
co-rotation enforcing charge.
The condition holds only within the light cylinder.
The direction of particles in the lab frame is given by
\begin{equation}
\boldsymbol {\beta}_0 = \beta'_{\parallel}\hat{\bf B} +
 \boldsymbol{\Omega}\times 
{\bf r}/c,
\label{2.4}
\end{equation}
where
\begin{equation}
\beta'_{\parallel} = -\hat{\bf B}\cdot(\boldsymbol{\Omega}\times{\bf
 r}/c) + 
 \{[\hat{\bf B}\cdot(\boldsymbol{\Omega}\times{\bf r}/c)]^2 -
 (\boldsymbol{\Omega} 
\times{\bf r}/c)^2 +1\}^{1/2},
\label{2.5}
\end{equation}
and $\hat{\bf B}$ is the unit vector along the magnetic field in lab frame.
The particle velocity is highly relativistic, so we have made the
approximation 
$|\boldsymbol{\beta}_0|\rightarrow 1$ in eq.~(\ref{2.5}).
Thus the direction of radiation 
emitted tangential to the particle velocity 
vector is given by $\boldsymbol{\beta}_0$ in eq.~(\ref{2.4}).
This direction is related to the periodic pulse phase. 
The observed phase $\phi$ is the sum of the azimuthal angle 
$\phi_{em} $ at the emission point and
the relativistic time delay \citep{RY95}:
\begin{equation}
\phi = -\phi_{em}
 -\frac{{\bf r}_{em}\cdot \hat{\boldsymbol{\beta}_0} }{ R_{LC}},
\label{2.6}
\end{equation}
where ${\bf r}_{ em}$ is the emission point, and  
$R_{LC}$ the light cylinder radius.

A certain mechanism is needed to fix the lower boundary of the 
particle acceleration region. 
In most works, including TCS08,
the lower boundary is 
chosen as the surface of the last-open
field lines of a rotating dipole in a vacuum.
The field lines are calculated by eq.~(\ref{2.2})
from the neutron star surface, 
and the last-open ones are defined as being just tangential to the
light cylinder and they form a magnetic surface from the polar cap.
The numerical procedure is described by \citet{CRZ00} in detail and 
we follow it.
In the outer gap model, if particle acceleration occurs in
 an open zone, the curvature radiation from the accelerated particles
 forms a narrow cone along the magnetic field lines in a frame rotating
 with angular velocity $\Omega$.
These $\gamma$-ray photons are converted by colliding X-ray photons
to $e^{\pm}$ pairs, which tend to
screen the accelerating electric field. However, there is no supply 
of pairs on the last-open field lines and hence no screening of
the electric field, since the $\gamma$-ray photons are emitted 
only toward higher altitudes above the last-open field lines \citep{CHR86}.
The `real' last-open field lines
may be different from ones in a vacuum \citep{R96, KO09}. 
 We therefore take into account this possible deviation of the boundary.
 We assume that dipole magnetic field is an approximation within the light 
 cylinder and use eq.~(\ref{2.2}) as the global magnetic field structure.
 Even if the overall structure is not different so much, critical value 
 between open and closed field lines is very sensitive to the boundary value 
 at the surface.
 The "real" last-open field lines do not in general agree with those in 
 vacuum field.
 Thus we introduce a parameter, altitude of the emission region as a 
 correction factor in order to take into account the deviation of boundary 
 from the vacuum field.
In our model this parameter specifies the range of the
emission region which is located above or below the last-open field 
lines within the light cylinder radius. 
Each different field line originating from the magnetic polar region
is parameterized by magnetic colatitudes
$\theta_m  $ and azimuthal angles $\phi_m$.
Following \citet{CRZ00},
we define open volume coordinates on the polar cap, 
($r_{ov}$, $\phi_m$), 
where $r_{ov}\equiv \theta_m/\theta^{pc, 0}_m(\phi_m)$. 
The function $\theta^{pc, 0}_m$ is the magnetic
colatitude of the conventional polar cap angle and
generally depends on the magnetic azimuth $\phi_m$.
The parameter $r_{ov}$ corresponds to the altitude of the emission region:
The last-open field lines of a rotating dipole in a vacuum
correspond to $r_{ov}=1$, whereas those for
higher altitudes have $r_{ov}<1$. 
Following \citet{TC09}, 
the maximum value is chosen as $r_{ov}= 1.36^{1/2}$, which 
corresponds to the polar cap angle 
$\theta^{pc}_m\sim 1.36^{1/2}\theta^{pc, 0}_m$, 
obtained in the force-free limit 
by \citet{CKF99}.
We found that no significant caustics are formed in the sky map, 
even if the maximum value of $r_{ov}$ is increased.

We assume that the radiation of different energy bands
is emitted from different field lines characterized by altitude.
The field line relevant to the $\gamma$- and X-rays is
approximated as being the same one. 
The direction of the emission is tangential to the lines, and
inward and outward directions are possible.
Both location and direction affect
the light curve profile of the energy bands.
Following the model by TCS08,
the $\gamma$-ray radiation above 100 MeV is emitted by particles moving 
in an outward direction, whereas radiation at 
lower energy bands is 
emitted by those moving in both outward and inward directions.
We use two conditions to constrain the emission region. 
First condition is the radial extension of the emission region. 
The outward emission is restricted to radial distances $r_n < R_{LC}$, and the
inward one is restricted to $r_s < r < \min (3r_n, R_{LC})$. The outer
boundary $3r_n$ for inward emission comes from the results of dynamic model
(TCS08), in which very few ingoing pairs are produced beyond the radial
distance $r > 3r_n$. 
Second condition is the azimuthal extension of the emission region. We use the
magnetic azimuthal angle of the footprint of field line (i.e., the point where
magnetic field line penetrates the neutron star surface) to characterize the
field line for given $r_{ov}$. Radial distance to the null charge surface on
the field lines significantly depends on the magnetic azimuthal angle. In the
outer gap model, most of the pairs are created around the null surface
(TCS08). We expect that the gap activity is related to the distance to the
null surface. Although the current density should be determined by global
conditions, there is no study of the three-dimensional magnetosphere of an
inclined rotator. In this paper, we assume that the field lines of both
outward and inward emission are active only if the radial distance to null
surface $r_n$ is shorter than $R_{LC}$. The azimuthal constraint is
automatically satisfied for outward emission because the radial extension
gives $r_n < R_{LC}$. However, for inward emission the condition becomes
strong. The radial extension $r_s < r < \min(3r_n, R_{LC})$ allows for the
regions $r < R_{LC}$ on the field lines with $r_n > R_{LC}$. They are not
active, so that the corresponding regions should be excluded. 
The critical value $3r_n$ was obtained by fitting to Vela pulsar (TCS08). 
It is not straightforward to apply it to other sources. The mean free path
$\lambda(r)$ of the pair creation process between the $\gamma$-ray and thermal
X-ray emissions from the stellar surface is estimated as $\lambda (r) \sim 5.6
P^{13/21}(B_s/10^{12}G)^{-2/7} r$ \citep{Ta08}. 
The value at the null pointis $\lambda(r_n)$ found to be in a range of $(2$-$3)r_n$ for our
samples. 
Our light curves especially
peak positions are not changed even by adopting $2r_n$ as the outer boundary
for inward emission.

Spatial distribution of the emissivity is approximated by
the step function-type, but the peak positions weakly
depend on the detailed emissivity distribution.

We assume that the overall structure of the light curve 
comes not from the emissivity distribution,  but 
from a bunch of many field lines in the observation, 
that is, caustics.
The appearance of caustics strongly depends on 
the observational viewing angle $\xi$ and the 
intensity distribution.
In this paper, we focus on the peak phases of the light curve, 
so we adopt a simple, uniform emissivity 
along all magnetic field lines,
which is independent of both the magnetic azimuthal angle 
$\phi_{m}$ and the altitude $r_{ov}$.
The fitting does not completely reproduce 
the observations so, in Section 4, a simple improvement to 
the emissivity distribution
is considered which leads to a much better fit.

We now explain our fitting method.
For fixed inclination angle $\alpha $ and viewing angle $\xi$,
the light curve as a function of phase $\phi$
depends only on the altitude $r_{ov}$.
The intensity is calculated
in the range $ r_{ov}< 1.36^{1/2}$ with a bin width of 0.02.
There are no significant caustics for large $ r_{ov}$. 
%

In the observed light curve, the reference phase $\phi=0$ is 
assumed to be located at the radio emission peak maximum 
in most studies (e.g., \cite{Abcat}).
However in the model light curve, the conventional reference phase $\phi=0$ 
occurs when the magnetic axis, spin axis and Earth line of sight lie 
all in the same plane.
These two reference phases do not agree with each other since
it is generally assumed that radio emissions arise at non-zero
altitude in most empirical studies. 
Following \citet{RW10}, we allow 
a shift by $-0.1 \le \delta\phi \le 0.1$ in 
the model reference rotation phase.
This degree of freedom does not significantly 
affect the determination of the altitude parameter $r_{ov}$,
because we use the peak separation in the observed
$\gamma$-ray and X-ray light curves which are emitted at 
the same $r_{ov}$.
For the sources showing a double-peak pulse shape in the
observed $\gamma$-ray light curve, 
we use the peak separation.
For those showing a single-peak, 
we use the phase separation between 
the $\gamma$-ray peak and one of X-ray peaks.
This is the benefit of considering $\gamma$-ray and X-ray light curves
simultaneously.
Subsequently, we look for the altitude of the UV/optical emission 
region using the $\gamma$-ray upper limit of $r_{ov}$.

\section{RESULTS}

In this section, we compare our model with pulse profiles observed at
multiple wavelengths for seven pulsars.
The sources are chosen using two criteria.
One is that non-thermal pulses are detected in addition to the $\gamma$-ray
and radio bands.
Our concern is to explore whether or not the emission region for
different energy bands is explained by the TCS08 model.
The second criterion is that the geometrical parameters,
$\alpha$ and $\xi$ are observationally constrained by the relativistic 
Doppler-boosted X-ray pulsar
wind nebula (PWN) \citep{NR08} or radio polarization data (e.g. 
\citet{LM88}).
The torus fitting method constrains the viewing angle $\xi$ only.
A small allowed range of $|\alpha-\xi|\leq 10^{\circ}$  is assumed 
for samples in which only $\xi$ is constrained due to the fact that radio 
emission from the pulsar polar region is detected. 
The geometrical parameters for the pulsars are listed in Table 1. 
We use these values, although there are some uncertainties in them.
The results are summarized in Figs.~\ref{fig:3.1} and ~\ref{fig:2.1}. 
  Fig.~\ref{fig:3.1} shows the intensity map for outward (upper panel) and
  inward (lower panel) emission as a function of the altitude of the emission
  region $r_{ov}$ and rotational phase. The upper and lower panels in
  Fig.~\ref{fig:2.1} are their pulse profiles for outward and inward emission
  in 
$\gamma$-ray and X-ray emission regions.

\subsection{Vela pulsar (PSR J0835-4510)}

We start with the Vela pulsar, which has been well studied to
test the validity of our simple model.
TCS08 considered
this source, but they used geometric parameters that are slightly 
different from ours.

\begin{figure}
 \begin{center}
  \includegraphics[width=160mm]{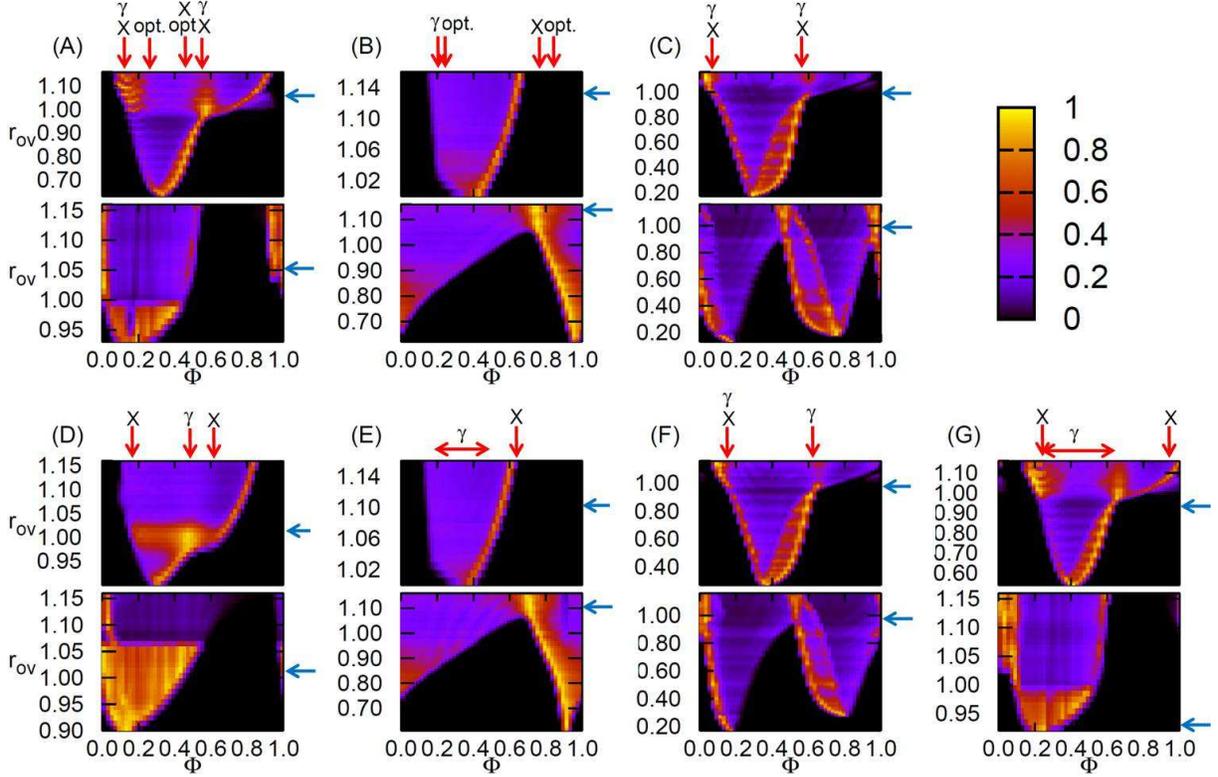}
 \end{center}
 \caption{The intensity maps for seven pulsars. 
 Upper left to right: PSRs J0835-4510 (A), J0659+1414 (B) and
 J0205+6449 (C).
 Lower left to right: PSRs J2229+6114 (D), J1420-6048 (E),
 J2021+3651 (F) and
 J1057-5226 (G).
 In each sample, upper panel is outward emission and lower is inward
 emission. 
 The blue horizontal arrows show best fit values of $r_{ov}$ for
   $\gamma$-ray and  X-ray emission regions.
 The red vertical arrows show the phase of peaks.
 The red horizontal arrows in (E) and (G) show the phase range of broad
 peaks.} 
 \label{fig:3.1}
\end{figure}

\begin{figure}
 \begin{center}
  \includegraphics[width=100mm, angle=270]{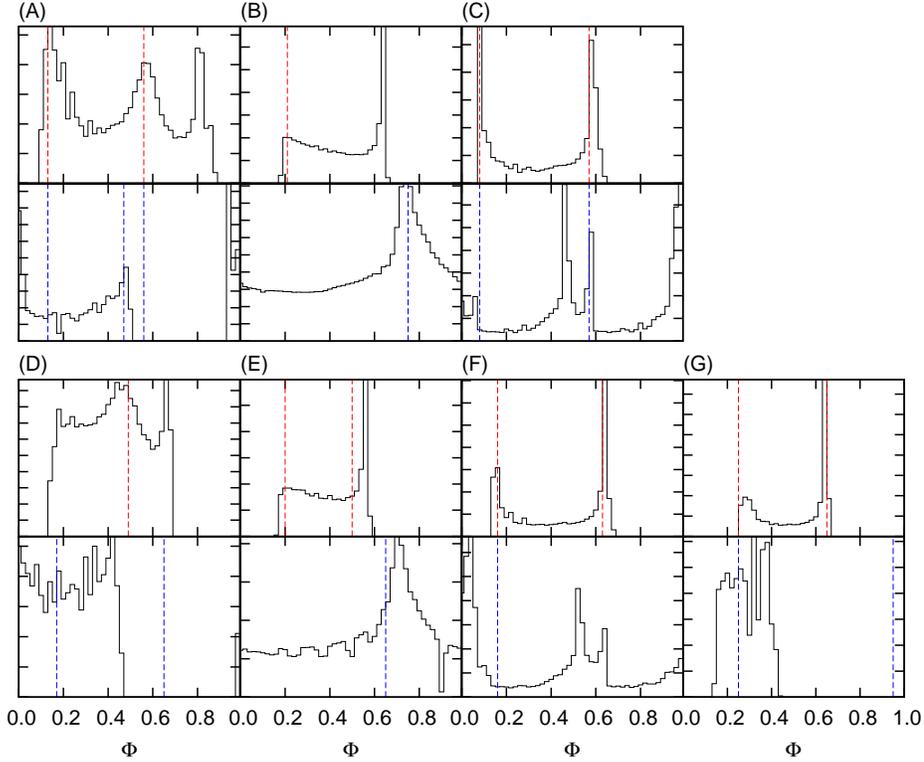}
 \end{center}
 \caption{The calculated light curves for $\gamma$- and X-ray emission region. 
 Upper left to right: PSRs
   J0835-4510 (A), J0659+1414 (B) and  J0205+6449 (C).
 Lower left to right: PSRs J2229+6114 (D), J1420-6048 (E),
 J2021+3651 (F) and
 J1057-5226 (G).
 In each sample, upper panel is outward emission and lower is inward
 emission.
 The vertical axis is in arbitrary units.
 The red and blue short-dashed vertical lines show the phase of $\gamma$-ray
 and X-ray 
 peaks as in Fig  ~\ref{fig:3.1}. 
 The peaks of PSRs J1420-6048 and J1057-5226 are so broad that the phase range
 is within two red vertical lines.} 
 \label{fig:2.1}
\end{figure}

Pulse profiles have been detected in optical to $\gamma$-ray
bands. The observed pulse profile in $\gamma$-ray band 
by {\it Fermi} \citep{Abvela}
shows a prominent double-peak structure and bridge
emission between the two peaks.
The first and second peaks are located at the phases $\phi\sim0.13$ and
$\phi\sim0.56$, respectively, and the separation 
is $\Delta\phi=0.43$.
We show the intensity map for outward emission as a function of the
altitude of the emission region and rotation phase 
in upper panel of Fig.~\ref{fig:3.1}(A). 
The emission altitude producing a peak separation 
$\Delta\phi=0.43$ is $r_{ov}\sim1.05$-1.06.

The X-ray data from {\it RXTE} \citep{Ha02}
also shows a double-peak structure but the second peak broadens 
toward early phase. 
The calculated intensity map is shown 
for outward and inward emissions
in the upper and lower panels of Fig.~\ref{fig:3.1}(A).
The main double peaks are located at the same phases as those 
in the $\gamma$-ray band, so they are interpreted as being 
formed by outward emission. 
The broad component before the second peak at $\phi \sim 0.47$
is associated with the caustic formed by the inward emission, 
as shown in the lower panel.
We attempted a fit without the inward emission, but found that
the inward emission is needed 
in order to reproduce the observed X-ray pulse profile.
The necessity of inward emission was discussed in TCS08. 
Thus, the peak positions of $\gamma$- and X-ray pulses 
can be explained with the same value of $r_{ov}\sim1.05$-1.06.
The contour map, however, shows a minor peak at $\phi\sim0.8$
formed by the outward emission. The peak was not observed in 
the $\gamma$-ray band.

We compare our model with the UV data of \citet{RKP05}
and optical data of \citet{G98}.
The pulse profiles in both bands are very similar, that is, they have 
a double-peak structure at the same phases. The
peak phases however differ from those of the $\gamma$- and X-ray bands. 
The first peak of the UV/optical bands shifts to a 
later phase $\phi\sim0.27$ and the second
peak shifts to an earlier phase $\phi\sim0.46$, so that
the peak separation becomes smaller. 
It can be seen from Fig.~\ref{fig:3.1}(A) 
that such a double-peak structure corresponds to
$r_{ov}\sim0.65$-0.80 for the outward emission.
The corresponding inward emission 
cannot be detected since its observable range is  
$r_{ov}  \ge 0.9$, as shown in the lower panel.
The choice of outward emission with $r_{ov}\sim0.65$-0.80 
is also supported by the fact that 
the second peak at $\phi\sim0.46$
in the UV/optical ranges is sharper than the first one at 
$\phi\sim 0.27$, because of their different dependence on
$r_{ov}$.
Thus, we have reproduced the pulse profiles 
of optical to $\gamma$-ray bands by the caustics model
without any detailed assumptions about emissivity. 
From the fitting model, we found three conditions for the
emission region. 
(1) The UV/optical emission region is located at an altitude
above the $\gamma$- and X-ray emission region 
of $r_{ov}\sim1.05$-1.06.
(2) There is a separation of altitude between 
the X-ray and optically dominant emission regions.
(3) The UV/optical emission range, $\Delta r_{ov}\sim0.15$,
is broader than that of $\gamma$/X-ray emission regions,
$\Delta r_{ov}\sim0.02$. 

\subsection{PSR J0659+1414}

The pulsar PSR J0659+1414 has also been observed
in the $\gamma$-ray to optical bands.
We use the $\gamma$-ray data from {\it Fermi} \citep{We10}, 
X-ray data from {\it XMM-Newton} \citep{DL05}, 
UV data from \citet{Sh05} and 
optical data from \citet{Ke03}.
The X-ray data is a combination of thermal (blackbody) and
non-thermal (power-law) emissions and is consistent with a cooling
middle-aged neutron star (e.g., \citet{BT97}).
At soft X-rays, the pulse fraction is low and the pulsations are
sinusoidal, as is typical for thermal emissions from the surface of a neutron
star with non-uniform temperature distribution. At higher energies
($>$1.5keV), where the non-thermal component dominates, the pulsed
fraction increases and the profile becomes single peaked.
We, therefore, consider the pulse profiles of hard X-rays ($>$1.5keV) 
only. 

The pulse profile in the $\gamma$-ray band 
shows a relatively broad single peak, which lags the radio maximum 
peak by $\phi\sim$0.2 in phase.
The peak in the non-thermal X-ray pulse is at
$\phi\sim0.7$-0.8, which is different from 
the phase of the $\gamma$-ray peak.
This phase difference cannot be ignored, although
the peaks in the $\gamma$- and X-ray data are rather broad, and
hence the difference may be diminished somewhat by including the phase error. 
We interpret these pulse profiles as being
emissions at different phases and different directions:
the peak of $\gamma$-ray is formed by outward emission, whereas
that of X-ray is formed by inward emission.
The intensity maps are given in the upper panel (outward emission) 
and lower panel (inward emission)
of Fig.~\ref{fig:3.1}(B), respectively.
From this figure, we see that 
a peak separation $\Delta\phi=0.55$ 
between the $\gamma$-ray and X-ray data corresponds to an emission altitude of
$r_{ov}\sim1.13$-1.14 
by shifting the reference phase by $\delta\phi=0.06$ 
to an earlier phase.
The emission altitude cannot be fixed without the X-ray data:
a shift of peak phase is allowed, so $r_{ov}$ is
unknown.
This ambiguity is removed by considering multi-wavelength light curves.
From the intensity map, we expect another very sharp caustic 
at $\phi\sim0.65$, but
there is no counterpart in the $\gamma$-ray observations
(top panel of Fig.~4 in \citet{We10}).
We discuss this missing peak in Section 4.2.
In the lower right panel of Fig.~4 in \citet{We10}, 
the light curve for inward emission is given.
The X-ray data may be a combination of 
inward and outward emissions, but the X-ray profile observed by \cite{DL05}
is similar to that of inward emission only.
This means phenomenologically that 
outward emission of X-rays is weak in this source.

The pulse profiles in the UV \citep{Sh05} and optical bands \citep{Ke03}
are almost the same shape and have a clear double-peak structure 
unlike the single peak in the $\gamma$- and X-ray bands.
The first peak at $\phi\sim0.02$ is later than 
the $\gamma$-ray peak 
and the second peak at 
$\phi\sim0.10$ is later phase than the X-ray peak.
From the upper panel of Fig.~\ref{fig:3.1}(B), we see that
the observed first peak can be reproduced by 
outward emission at an altitude $r_{ov}<1.10$.
Here we have a weak condition because the peak
position depends only weakly on $r_{ov}$. 
For the second peak, the inward emission forms caustics 
for $0.90<r_{ov}<1.04$.
Thus, the altitude of the emission region in the UV/optical bands
is identified as $r_{ov}\sim0.90$-1.04, where 
the lower limit is set by a coarse bin of the phase
in the observational data.

\citet{Ke03} have already investigated the 
multi-wavelength light curve of this pulsar 
using a similar method to ours, but could not explain the profile
using geometrical parameters which are consistent with radio
polarization data \citep{EW01}.
The reason for this is that the lower boundary of
the emission region was chosen as 
the last-open field lines in the vacuum dipole field, that is,
$r_{ov}=1.0$.
In our analysis, by allowing $r_{ov}\geq1.0$, the phase of peaks can be
  successfully fitted by using observed geometrical parameters.
This suggests that the actual lower boundary of the gap is 
slightly different from the last-open field lines in a vacuum dipole.

\subsection{PSR J0205+6449}

The $\gamma$-ray pulse profile observed by {\it Fermi} \citep{Ab58}
shows a double-peak structure.
The first peak is offset from the radio peak by $\phi\sim0.08$, and
the second is at $\phi\sim0.57$.
The separation is $\Delta\phi=0.49$. 
The X-ray data from {\it RXTE} \citep{Li09} are consistent with the
results from {\it XMM-Newton} \citep{Ku10}.
The spectral shape can be fitted by a power law such 
that most of the emission is non-thermal and the thermal component is
constrained by the upper limit \citep{Ku10}.
The observed X-ray pulse profile shows two peaks aligned in phase
over a wide energy range of $\sim$0.5-270 keV,
and is also very similar to that of the $\gamma$-ray band.

We show the intensity maps for outward and inward emission in 
the upper and lower panels of Fig.~\ref{fig:3.1}(C), respectively.
As seen in the upper panel, the emission altitude for the
double-peak with $\Delta\phi=0.49$ is $r_{ov}\sim0.97$-0.98.
A shift of the reference phase $\delta\phi$ is not necessary 
in this source.
As argued in TCS08,
outward emission dominates in the light curve
for a young pulsar with a strong non-thermal X-ray component,
like the Crab pulsar.
PSR J0205+6449 is the youngest pulsar in our sample (characteristic age
$\tau_c\sim 5\times 10^3$ yr) and shows rather strong non-thermal
radiation in the X-ray band.
Thus, it is likely that 
only outward omission contributes to the observed X-ray light curve
in this pulsar. 

\subsection{PSR J2229+6114}

The light curve for $\gamma$-rays 
observed by {\it Fermi} \citep{Ablv}
shows an asymmetric single peak at $\phi\sim0.49$.
The tail extends down to $\phi\sim0.2$.
The peak position depends slightly on the energy range above 100 MeV, 
but the amount of shift is only $\sim0.04$. 
The X-ray pulse profile observed by {\it XMM-Newton} \citep{Ablv}
shows a double-peak structure.
No peak is seen at the $\gamma$-ray peak phase.
The separation between first X-ray peak and the $\gamma$-ray peak is
$\Delta\phi\sim0.32$, and the separation
between the second X-ray peak and the $\gamma$-ray peak 
is $\Delta\phi\sim0.14$.

The intensity map for outward emission is shown in the upper
panel of Fig.~\ref{fig:3.1}(D).
We consider the formation of the $\gamma$-ray peak and 
two X-ray peaks as being due to outward emission only.
Such a solution is possible by choosing 
an emission altitude of $r_{ov}\sim1.01$-1.02
with a small phase shift $\delta\phi=0.03$.
The intensity map for inward emission shows
a sudden decrease in the number counts for $r_{ov}<1.06$.
The peak becomes broad and hence the contribution of the inward 
emission is not very important.   
We show the light curve of inward and outward emission
for $r_{ov}\sim1.01$-1.02
in Fig.~ \ref{fig:3.1}(D).
The outward emission curve is very similar to 
the observations in the $\gamma$- and X-ray bands.
In this pulsar, we need to use all the light curves
simultaneously in order to determine the range of $r_{ov}$.
Since emissions with smaller values of $r_{ov}$ are not seen, 
if $\gamma$-ray and optical emission regions
are separated $\Delta r_{ov}>0.10$, similar to the Vela pulsar and
PSR J0659+1414, we predict that an optical pulse
profile cannot be observed or will be only very weakly detected.
This is consistent with the fact there have been no
reports of the detection of a pulse in the lower energy band for this pulsar.


\subsection{PSR J1420-6048}

The $\gamma$-ray light curve from {\it Fermi} \citep{We10}
shows a broad peak at $\phi\sim0.2$-0.5.
This peak may consist of two components, but it is not clear in
the current photon statistics.
The X-ray pulse profile from {\it ASCA} 
is detected weakly at a marginal level, and shows a broad peak at 
$\phi\sim0.6$-0.7, which is different from the $\gamma$-ray peak \citep{Ro01, B09}.
Recently, in the table of \citet{MDC11} they list this object as
  non-thermal dominated source in X-ray. The pulsed X-ray
  profile is likely to originate from the non-thermal component.

Since the light curve of this pulsar and its geometrical parameters
are similar to those of PSR J0659+1414,
we adopt the same interpretation. 
That is, the $\gamma$-ray peak is formed by outward emission, 
whereas the X-ray peak is formed by inward emission.
From the intensity map in Fig.~\ref{fig:3.1}(E), 
an emission altitude of $r_{ov}\sim1.10$-1.11 
corresponds to one broad peak at 
$\phi\sim0.2-0.5$ by outward emission 
and another at $\phi\sim0.6-0.7$ by inward emission. 
Here a small shift $\delta\phi=0.10$ toward earlier phase
is used.
Since there are similarities in both the $\gamma$- and X-ray light curves 
and the geometrical parameters between this pulsar and
PSR J0659+1414, 
we expect a similar double-peak pulse profile in the optical band, 
if it is detected.

\subsection{PSR J2021+3651}

Observations in the $\gamma$-ray band have been obtained 
by {\it Fermi} \citep{Ab20} and
{\it AGILE} \citep{Ha08}. 
The observed light curve shows a sharp double-peak structure.
The first peak is offset from the radio peak by $\phi\sim0.16$ and 
the two peaks are separated by $\Delta\phi\sim0.47$. 
The X-ray light curve in \citet{Ab20} shows
a relatively sharp peak associated with first peak in the $\gamma$-ray
light curve albeit with weak photon statistics.
In this paper, we assume that at least the first peak is non-thermal in
origin.
The possible contribution of non-thermal X-ray emissions is also
discussed in \citet{He04} and \citet{VRN08}.
We expect that this assumption will be tested by phase-resolved spectra
from future observations.

As seen in the upper panel of Fig.~\ref{fig:3.1}(F), the emission
altitude is $r_{ov}\sim0.97$-0.98,
for which there is a $\gamma$-ray double-peak profile with separation
$\Delta\phi=0.47$ and a relative shift $\delta \phi=0.06$ toward later phase.
The peak of non-thermal emission at $\phi=0.15$-0.20 in the X-ray light curve
is found to be formed by outward emission only.
The relatively weak second peak in the X-ray band is consistent
with the case of PSR J0205+6449.
Thus, the three model parameters for this pulsar and
PSR J0205+6449 are very similar, as shown in Table 1.

\subsection{PSR J1057-5226}

The $\gamma$-ray light curve from {\it Fermi} \citep{Ab3} 
shows a broad peak at $\phi\sim0.25$-0.65.
This probably consists of two components, but it is not clear.
\citet{DL05} extract only a power-law component
of the X-ray light curve and their Fig.~13 shows
a two peaks at $\phi\sim0.2$-0.3 
and $\phi\sim0.9$-1.0, although the data are very coarse.
We regard the light curve as being produced by a non-thermal X-ray component. 

The observed light curve may be regarded 
either as a broad peak consisting of weak peaks and a relatively bright
bridge emission or as a result of the range of the emission region widening  
towards lower altitudes.
In the latter interpretation the fitted $r_{ov}$ is only a lower limit.
We thus focus on the width of the $\gamma$-ray peak.
From the upper panel of Fig.~\ref{fig:3.1}(G), the emission altitude 
is $r_{ov}\sim0.93$-0.94
Even if we assume a double-peak structure with the first
peak at $\phi\sim0.31$ and the second peak at $\phi\sim0.59$ following
\citet{Ab3}, we have $r_{ov}\sim$0.90-0.91, which is very
similar to the value obtained above.
Thus we have $r_{ov}\sim$0.90-0.95 in either case.
The phase shift is $\delta\phi=0.10$ in this pulsar.
The first peak in the X-ray light curve
is formed by outward emission, but the second one 
cannot be produced for the same altitude. 
This may be a drawback to our model, but 
the present X-ray data are coarse and a 
much more precise non-thermal X-ray light curve is needed.

\section{DISCUSSION}

\subsection{Statistical properties of the emission region}

In the previous section, 
we have shown that 
   the peak phases of seven pulsars emitting $\gamma$- and X-rays
  can be successfully fitted 
using the TCS08 outer gap model,
in which both $\gamma$-rays and X-rays originate from the same magnetic 
field line characterized by an altitude $r_{ov}$.
The parameter $r_{ov}>1$ is needed in the light curve 
fitting for some sources.
Moreover, the inclusion of inward emission for X-rays causes 
a variety of pulse profiles in both bands.
The parameter $r_{ov}$ could not be determined solely using $\gamma$-ray 
data for a single $\gamma$-ray peak pulsar.
But, by considering the X-ray light curve, the parameter
is uniquely determined for PSRs J0659+1414, J2229+6114
and J1420-6048.

\begin{figure}
 \begin{center}
  \includegraphics[width=100mm, angle=270]{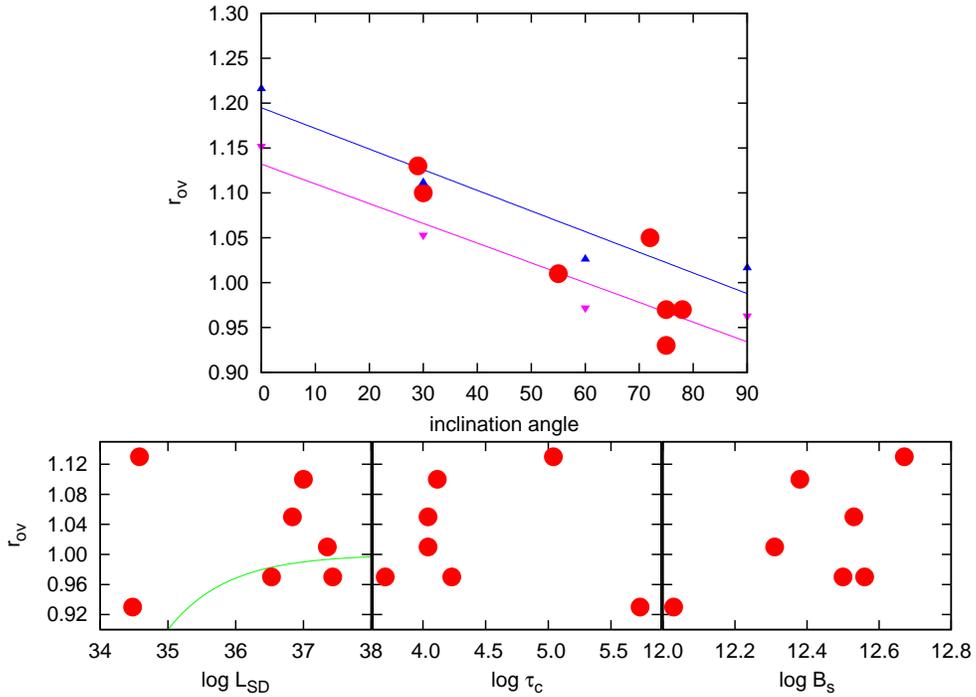}
 \end{center}
 \caption{The relation between $r_{ov}$ and inclination angle (upper),
 spin-down luminosity (lower left), characteristic age
 (lower middle) and surface magnetic field (lower right).
 The altitudes corresponding to the separatrix layer model are shown as purple
 downward and blue upward triangles in the upper panel. The two lines are
 linear 
 fitting lines for the separatrix layer model. The light green curve in the
 lower left panel shows the relation
 $(1-r_{ov})=(10^{33}ergs^{-1}/L_{SD})^{1/2}$.} 
 \label{fig:4.2}
\end{figure}

It is worthwhile to explore the general dependence of 
the altitude $r_{ov}$ on other characteristics if any,
although there may not be enough data for a proper statistical analysis.
In Fig.~\ref{fig:4.2},
$r_{ov}$ is plotted as a function of inclination angle $\alpha$, 
spin-down luminosity $L_{sd}$,
characteristic age $\tau_c$ and surface dipole magnetic field
$B_s$.
We found that there is a significant 
correlation between $r_{ov}$ and the inclination angle $\alpha$ only;
the relations of $r_{ov}$ with the other parameters are very weak.   
This correlation suggests that the deviation 
from a vacuum rotating dipole field is large
for small inclination angle.
It is very interesting to compare 
this result with that in a force-free magnetosphere. 
\citet{BS10b} proposed
that the separatrix layer at an altitude of 
0.90-0.95 times the height of the last-open field 
line is relevant to emissions in a three-dimensional
inclined force-free magnetosphere.
This altitude, which is not exactly symmetric with respect to
the magnetic azimuthal angle $\phi_m$,  
but can be approximated by the value at $\phi_m=0$,
is plotted in Fig.~\ref{fig:4.2}
as purple downward and blue upward triangles.
Two linear fitting lines are also shown. 
The altitude $r_{ov}$ decreases with
the inclination angle $\alpha$ in both  
our model and the separatrix layer model of a force-free magnetosphere.
However, the emission region in the separatrix layer model 
extends even outside the light-cylinder, whereas
ours is well localized around null points.
Accounting for this difference may be important for further 
improvement of the model
of the emission region based on a force-free magnetosphere.


The thickness of the gap region, $w$, is not known, but it is sometimes 
assumed to decrease with the spin-down luminosity $ L_{SD}$
\citep{Wa09, RW10}.
We have $w=1-r_{ov}$, if
the lower boundary of the gap is fixed 
as the last-open field line in the vacuum dipole field.
This assumption is tested 
in the lower left panel of Fig.~\ref{fig:4.2}, in which
the relation 
$(1-r_{ov}) \approx (L_{SD} / 10^{33}erg s^{-1})^{-1/2}$ 
is plotted as a light green curve.
(The curve is not fitted to the data points.)
This suggests that the assumption of maximum altitude, 
$r_{ov}=1.0$, is not a good one.
This discovery affects expected number of the $\gamma$-ray pulsars 
in the observation.
From geometrical reason, the pulsed emission by caustics is 
limited to a certain range between inclination and viewing angles. 

\begin{figure}
 \begin{center}
  \includegraphics[width=100mm]{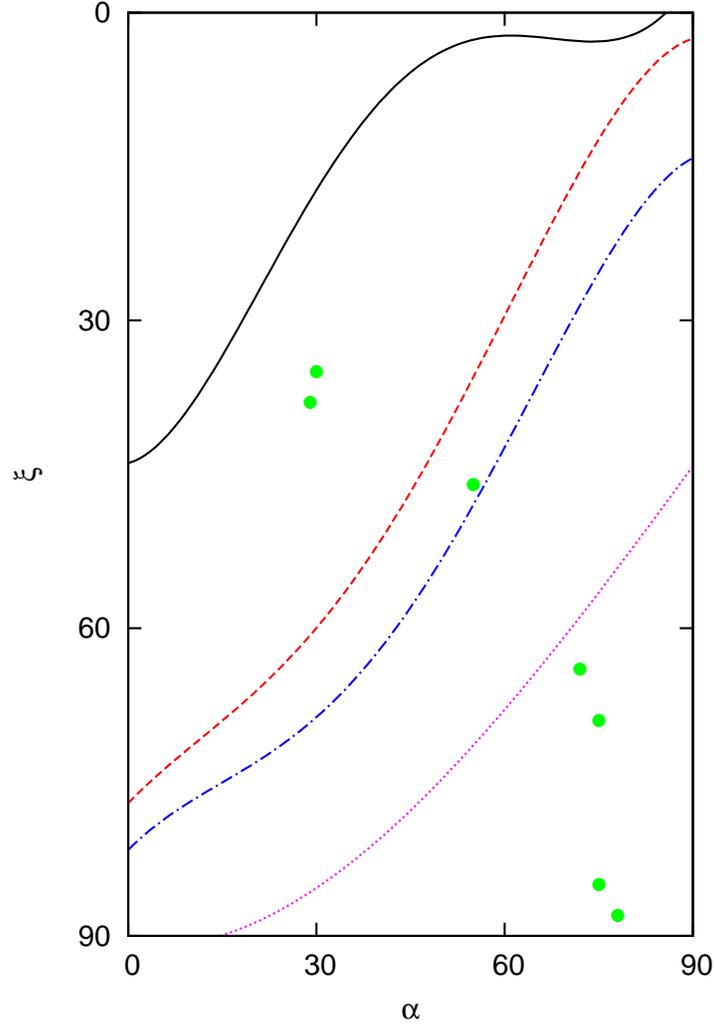}
 \end{center}
 \caption{The observable range for $\gamma$-ray pulsars in 
 the $\alpha$-$\xi$ plane for the outer gap model. Black solid curve shows 
 the boundary of observable pulsars using linear fitting line for the
   separatrix layer model. 
 Red dashed, blue dash-dotted and purple dotted curves show the boundary 
 with $r_{ov}=$0.95, 0.90 and 0.70, respectively. 
 Light-green curcles show the pulsars in Table 1.}
 \label{fig:4.3}
\end{figure}

\citet{RW10} showed the range of observable pulsars with $r_{ov}=$0.95, 0.90 
and 0.70 for outer gap model in their Fig. 16.
We recalculate it and show the result in Fig.~\ref{fig:4.3}.
The observable range of viewing angle $\xi$ is below the curves. 
Our finding in Fig.~\ref{fig:4.2} is that $r_{ov}$ is a function of the 
inclination angle, which is similar to that of the separatrix layer model. 
We also show the observable range by the empirical relation obtained 
in Fig.~\ref{fig:4.2} as black solid line, for which the altitude is
chosen as 0.925 times the height
of the last-open field line in force-free magnetosphere.
The figure shows that sources with low inclination and viewing angles 
become observable. 
For example, pulsar with the inclination angle $\alpha =30^{\circ}$ can 
be detected for $\xi > 60^{\circ}$ for $r_{ov}=$0.95, but for $\xi >30^{\circ}$. 
Thus expected number increases approximately twice for sources 
with the low inclination and viewing angles.

\subsection{The phenomenological limitation for emissivity}

\begin{figure}
 \begin{center}
  \includegraphics[width=100mm, angle=270]{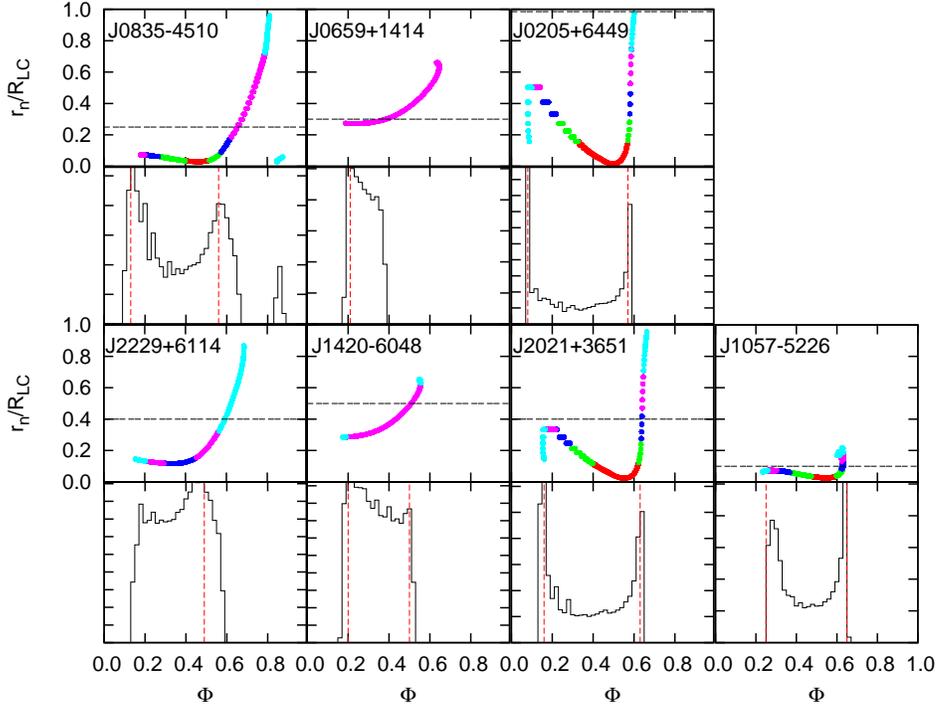}
 \end{center}
 \caption{The distribution of the radial distance to the null point of the
 field line on which observed photons are emitted (each upper panel) and
 the light curves 
 that are restricted by the azimuthal extension limit as a function of the
 rotation phase (each lower panel) for seven pulsars. 
 The color shows the radial distance to the emitting point as $0.0<r/R_{LC}<0.2$
 (red), $0.2<r/R_{LC}<0.4$ (light green), $0.4<r/R_{LC}<0.6$ (blue),
 $0.6<r/R_{LC}<0.8$ (purple), $0.8<r/R_{LC}<1.0$ (light blue).  
 The values $r_{n,lim}$ for each pulsar are shown as black long-dashed
 horizontal lines. 
 The red short-dashed vertical lines show the phases of the $\gamma$-ray
 peaks.
 For PSRs J1420-6048 and J1057-5226, the vertical lines show the
 phase range of broad peaks.}
 \label{fig:4.1}
\end{figure}

The caustic model considered in Section 3
provides peak positions consistent with observation,
but there are also some additional, unseen peaks.
These are interpreted as being prohibited by some mechanism.
In this section, we consider an improvement to our model that
takes into account a very simple distribution for the emissivity.
Detectable $\gamma$-rays are radiated with large multiplicity
by the pair plasma in the gap region.
Therefore, the mean free path of a $\gamma$-ray photon 
should be less than light cylinder radius \citep{TC07}.
The pair creation mean free path is given by
$\lambda(r)\sim 5.6 P^{13/21}(B_s/10^{12}G)^{-2/7} r$
 \citep{Ta08} for an
assumed limiting distance to the null point $r_{n,lim}$.
The position of the null point of inclined pulsars, where
the accelerating electric field arises, 
depends significantly on magnetic azimuthal angle, so
the intensity of $\gamma$-ray emission also depends on the magnetic
azimuthal angle.
Therefore, active field lines should be limited in the azimuthal 
direction. 
By taking into account the azimuthal extensions
with $\lambda(r_{n,lim})\simlt 0.2-0.7R_{LC}$ listed in Table 1,
the fits of the resultant light curves, which
are shown in Fig.~\ref{fig:4.1}, become better. 
In the same figure, we also show 
the radial distance to the emission points 
of the observed photons against
the rotation phase.
Note that, for the Vela pulsar, the minor third peak at $\phi\sim0.8$
still remains even after the inclusion of the azimuthal extension limit.
The corresponding radial distance of emission points is
relatively large, so that the photon energy is expected to be soft.
The third peak is not observed in the GeV band, but may appear
in a much lower energy band.
At least, the minor third peak of the X-ray light curve 
appears to be associated with the same caustic.
We have also tried to improve the X-ray light curve with some other simple
assumptions, but have not had good results.
The reason for this is that there are many ways for
X-ray emitting particles to be created: 
via thermal, magnetospheric emissions and magnetic pair creation.
Therefore, the three-dimensional effect of the propagation
of $\gamma$-ray photons and soft X-ray photons is very important.
Without it we cannot successfully explain the light curve. 

\subsection{The location of the UV/optical emission region}

We explored the UV/optical emission region for Vela and PSR J0659+1414.
The results are qualitatively similar:
the altitude range 
of UV/optical emission, $\Delta r_{ov} \sim 0.15$, is 
broader than that for $\gamma$- and X-rays,
$\Delta r_{ov} \sim 0.02$;
and both emission regions are not continuous and 
connected, but are widely separated.
The separation 
may come from two competing mechanisms: a decrease of emissivity
and an increase of synchrotron intensity in the UV/optical bands with 
altitude. 

As discussed in TCS08, outward emission is generally
dominant in UV/optical emission, as
shown in their Fig.~4. 
The explanation is the following.
The number of created pairs is the main cause of the difference
between inward and outward emissions in the UV/optical bands, since 
the collision angles with magnetospheric X-rays are not different
for outgoing and ingoing $\gamma$-ray photons  
in the acceleration region.  
More outgoing $\gamma$-rays are emitted, and hence
more outgoing secondary pairs are produced.
Thus, the synchrotron emission for outgoing secondary pairs
produced by magnetic X-rays is brighter than that 
for the ingoing secondary pairs.
The observed flux strongly depends on the geometrical 
configuration. The outward emission in the UV/optical bands
may not point toward us even though the intrinsic emission is strong.
Our results show that the peaks in the Vela pulsar
can be explained by outward emission alone, 
while those in PSR J0659+1414 
require both inward and outward emissions.
Our result suggests that outward emission
is significantly suppressed in PSR J0659+1414,
to the level of the intrinsically weak inward emissions.
The stronger component is hidden,
because the observable altitude range is narrow, as shown 
in Fig.~\ref{fig:3.1}(B). 
This may explain the fact that  
UV/optical flux is smaller than the value extrapolated
from non-thermal X-rays, 
as seen Fig.~4 of \citet{Mi10}, whereas
the flux coincides with the extrapolation in the Vela pulsar.
This interpretation may be tested in PSR J1057-5226.
We also suggest that this difference is the reason why 
PSR J0659+1414, which has similar geometrical parameters 
to the Vela pulsar,
has an observable optical spectrum \citep{MPK10}, 
and the flux is slightly smaller than the extrapolation from 
non-thermal X-ray emission. 
The pulse profile has not yet been determined, but
the peaks should appear at a phase 0.3$<\phi<0.6$
and be due to outward emission.

\section{SUMMARY}

 We have calculated the light curves of emissions using the TCS08 
outer gap model
and compared them with observed multi-wavelength light curves.
We find that the model can successfully explain the 
peak positions of multi-wavelength light curves.
In order to determine the altitude of the emission region, 
the observed X-ray light curve is important, especially when there is
a single peak in the $\gamma$-ray light curve.

The fit of a light curve based on a simple emissivity distribution
can be improved by taking into account
the limitation of azimuthal extension in which a 
reasonable value of the $\gamma$-ray photon mean free path
is adopted. 
The resulting difference between model and observed $\gamma$-ray light curves 
becomes small; 
however, there may still be an unseen peak, such as the
minor third peak in Vela.   
The best-fit values of the altitude of the emission region
for PSRs J0659+1414 and J1420-6048,
suggest a deviation 
from the last-open field lines of a vacuum dipole field.
The real last-open field lines lie inside those of 
vacuum dipole field, $r_{ov}<1.0$.
This shift suggests that the lower boundary is
very similar to that of a force-free magnetosphere. 
We find that the altitude of the emission region is
correlated with inclination angle. 
This relationship is also very similar to that in a force-free magnetosphere.
The lower boundary of emission region has been assumed to  
$r_{ov}=1$ so far, but our model fits do not support it. 
This modification of the boundary of the magnetosphere suggests 
that the pulsars with low inclination and viewing angles are likely to be
detectable.
Thus the expected number in the future observation in the 
previous works(\citet{TWC11, WR11}) is underestimated for the sources  
with low inclination and viewing angles.
%

\section*{Acknowledgements}
The authors thank S. Shibata and J. Takata for much valuable discussion.
This work was supported in part by the Grant-in-Aid for Scientific 
Research from the Japan Society for Promotion of Science(S.K.) 
and from the Japanese Ministry of Education, Culture, Sports,
Science and Technology(Y.K.  No.21540271).


\clearpage

\begin{table}
\rotatebox{90}{\begin{minipage}{\textheight}
\begin{center}
\begin{tabular}{ccccccccccc}
\multicolumn{11}{c}{TABLE 1 Pulsar parameters} \\ \hline
Name & $\log (L_{SD})$ & $\tau_c$ & $\log (B_s)$ & $\alpha$ & $\xi$  &
 Reference & $r_{ov}$($\gamma$-, X-ray) &
 $r_{ov}$(UV/optical)&$r_{n,lim}$&$\lambda(r_{n,lim})$ \\  
 & (erg s$^{-1}$) & (kyr)& (G) &(degrees) &(degrees)&  &  &
			     &($R_{LC}$)&($R_{LC}$) \\
(1) & (2) & (3) & (4) & (5) & (6) & (7) & (8) & (9)&(10)&(11) \\ \hline
J0835-4510 & 36.84 & 11 &12.53&72 & 64 & 1,2 & 1.05-1.06 &0.65-0.80&0.25&0.23\\
J0659+1414 & 34.58 & 110 &12.67& 29 & 38 & 3 & 1.13-1.14 & 0.90-1.04&0.30&0.60\\ 
J0205+6449 & 37.43 &   5 &12.56& 78 & 88 & 2 & 0.97-0.98 & $\cdots$&1.00&0.71\\ 
J2229+6114 & 37.35 &  11 &12.31& 55 & 46 & 2 & 1.01-1.02 & $\cdots$&0.40&0.29\\ 
J1420-6048 & 37.00 &  13 &12.38& 30 & 35 & 6,7 & 1.10-1.11 & $\cdots$&0.50&0.42\\
J2021+3651 & 36.53 &  17 &12.50& 75 & 85 & 5 & 0.97-0.98 & $\cdots$&0.40&0.40\\
J1057-5226 & 34.48 & 540 &12.03& 75 & 69 & 4 & 0.93-0.94 & $\cdots$&0.10&0.20
 \\ \hline
\multicolumn{11}{l}{}%
\label{tab:1}
\end{tabular}
\end{center}
NOTES.-Col.(1):Pulsar name. 
Col.(2),(3),(4):The spindown luminosity, the
characteristic age and the strength of surface magnetic field, which we
		adopt in \citet{Abcat}.
Col.(5):The inclination angle.
Col.(6):The viewing angle. 
Col.(7):Reference for cols.(5) and (6). 
Col.(8):The emission altitude in the $\gamma$-ray band. 
Col.(9):The emission altitude in the optical/UV band.
Col.(10):Assumed limit of radial distance to null point.
Col.(11):Mean free path for $\gamma$-ray photons at the limit of
radial distance to the null point.
REFERENCES.- (1)\citet{Jo06} (2)\citet{NR08} (3)\citet{EW01} (4)\citet{WW09}
(5)\citet{VRN08} (6)\citet{We10} (7)\citet{WJ08} 
\end{minipage}}
\end{table}


\clearpage

\clearpage

\clearpage


\end{document}